# Variational Auto-Encoder Based Approximate Bayesian Computation Uncertian Inverse Method for Sheet Metal Forming Problem

Jiaquan Wang [a,b] Yang Zeng [a,b] Xinchao Jiang [a,b] Hu Wang[a,b,γ] Enying Li[c]
Guangyao Li[a,b]

[a] State Key Laboratory of Advanced Design and Manufacturing for Vehicle Body, Hunan University, Changsha, 410082, China

[b] Joint Center for Intelligent New Energy Vehicle, Shanghai, PR China, 201804

[c] College of Mechanical & Electrical Engineering, Central South University of Forestry and Teleology, Changsha, 41004, P.R. China

**Abstract** In this study, an image-assisted Approximate Bayesian Computation (ABC) parameter inverse method is proposed to identify the design parameters. In the proposed method, the images are mapped to a low-dimensional latent space by Variational Auto-Encoder (VAE), and the information loss is minimized by network training. Therefore, an effective trade-off between information loss and computational cost can be achieved by using the latent variables of VAE as summary statistics of ABC, which overcomes the difficulty of selecting summary statistics in the ABC. Besides, for some practical engineering problems, processing the images as objective function can effective show the response result. Meanwhile, the relationship between design parameters and the latent variables is constructed by Least Squares Support Vector Regression (LSSVR) surrogate model. With the well-constructed LSSVR model, the simulation coefficient vectors under given parameters will be determined effectively. Then, the parameters to be identified are determined by comparing the simulated and observed coefficient vectors in ABC. Finally, a sheet forming problem is investgated by the suggested method. The material parameters of the blank and the process parameters of the forming process are identified. Results show that the method is feasibility and effective for the identification of sheet forming parameters.

**Keywords**: Approximate Bayesian Computation; Variational Auto-Encoder; LSSVR; Sheet forming

**Acronym**

ABC: Approximate Bayesian Computation
ABC-MCMC: ABC Markov Chain Monte Carlo
ABC-PMC: ABC Population Monte Carlo
ABC-NPMC: ABC Non-parametric Population Monte Carlo
ABC-PRC: ABC Partial Rejection Control
ABC-SMC: ABC Sequence Monte Carlo
AdaGrad: Adaptive Gradient
Adam: Adaptive Moment Estimation optimizer
BN-Conv: Batch Normalized-Convolution
FE: Finite Element
FLD: Forming Limit Diagram
LHD: Latin Hypercube Design
LReLU: Leaky Rectified Linear Unit
LSSVR: Least Squares Support Vector Regression
LV-LSSVR: LSSVR model based on latent variable
ReLU: Rectified Linear Unit
RMSProp: Root Mean Square Prop
SSIM: Structural Similarity
STD: Standard Deviation
VAE: Variational Auto-Encoder

# 1 Introduction

At present, the hybrid numerical method [1-6] based on the combination of experiment and simulation is a commonly used inverse method. For the hybrid numerical method, which is based on deterministic identification, little attention is paid to the uncertainty in the reverse process. However, uncertainties generally exist in engineering practice [7]. The uncertainties usually arise from material constitutive model parameters, structural geometric parameters, boundary conditions, initial conditions, measurement information, cognitive judgment and calculation methods [8, 9]. Therefore, the deterministic reverse method is difficult to measure the credibility of the reverse result in the intricate practical engineering. Due to the existence of uncertainty, small fluctuations in the design parameters may lead to large deviations in the response space [10]. Therefore, the parameter identification of uncertainty is meaningful for analyzing the degree of influence of the inevitable uncertainty factors on the results. Meanwhile, more parameter information can be obtained by the uncertainty analysis [11].

In the existing uncertain analysis methods, Bayesian inference [12] is widely used to solve inverse problems due to its flexibility. Some investigations have shown the feasibility of Bayesian inference in the parameter identification of uncertainty [13-18]. Under the Bayesian framework, the probability distribution of unknown parameters can be obtained with prior information. However, the likelihood function of Bayesian framework is always difficult to obtain for complex engineering problems. In order to address this problem, Pritchard et al. proposed an Approximate Bayesian Computation (ABC) method that bypassed the intractable likelihood function [19]. In the ABC, the distance between simulated data and observed data is compared [20]. If the distance satisfies a small tolerance value, the corresponding parameter values of simulated data are retained, vice versa. Generally, generation of a set of simulated data with a very small distance from observed data with high data dimensions is extremely expensive and even prohibited. Therefore, the low-dimensional summary statistics are used to replace the high-dimensional data [21]. There are two requirements for the summary statistics. First, in order to improve the computational efficiency, the scale of the summary statistics should be as small as possible. Second, in order to minimize information loss, the features of the high-dimensional data should be extracted to the utmost. However, this trade-off is difficult to achieve. Therefore, a large number of researchers focused on the selection of the summary statistics. Many methods have been proposed, such as regularization approach [22], the best subset selection method [23], projection methods [24, 25], etc. More details can be found in Prangle's summaries [26].

In order to compromise information loss and computational efficiency, images are mapped to a low-dimensional latent space using Variational Auto-Encoder (VAE) in this study. The VAE was first proposed by Kingma et al [27]. In recent years, the VAE has already shown promise in extracting features and reconstructing samples of a complex model. Li et al. (2017) proposed a new unsupervised sentence saliency framework for multi-document summarization by modeling the observed sentences and their corresponding latent semantics with the VAE. The test results proved that the proposed framework could achieve a better performance [28]. Hjelm et al. (2016) demonstrated that VAE was a viable method for feature extraction with magnetic resonance imaging data by dimensionality reduction and reconstruction of the raw data [29]. Pekhovsky et al. (2017) investigated a new *i*-vector speaker recognition system based on VAE, in which the VAE played an important role for capturing the characteristics of complex input data distribution [30]. Eleftheriadis et al. (2016) combined the Gaussian process with the VAE to ordinal prediction of facial action units and the test has obtained satisfactory results [31]. For VAE, the image can be trained to obtain a set of feature latent variables which can fully reflect the image characteristics. Moreover, the dimensions of feature latent variables in VAE may be set as needed. Therefore, the latent variables in the VAE can be used as the summary statistics, which can obtain more model information at a smaller scale. Meanwhile, in order to conveniently use the feature latent variables in the VAE, Least Squares Support Vector Regression (LSSVR) surrogate model [32] is used to construct the relationship between design parameters and feature latent variables. Therefore, the computational efficiency can be greatly improved by an accurate LSSVR model. Finally, the parameters can be determined by comparing the difference between the observed and simulated coefficient vectors composed of characteristic latent variables.

In order to obtain the posterior distribution of the material parameters efficiently, it is also important to select a suitable sampling method for the ABC. Actually, there are many kinds of sampling methods for the ABC, and the ABC rejection sampling method is the simplest one [19]. However, the ABC-rejection sampling is not the efficient one due to its high rejection rate for a very small tolerance value. Therefore, Marjoram et al. [33] extended the Markov Chain Monte Carlo (MCMC) method into the ABC framework to increase the acceptance rate of samples. However, the ABC-MCMC sampling method is difficult to converge for complex problems [34]. Meanwhile, the ABC Partial Rejection Control (ABC-PRC) sampling method uses the idea of the particle filter to make up for the defects of ABC-MCMC sampling [35]. Beaumont et al. explained that the ABC-PRC sampling might cause a biased estimate of the posterior, and proposed the ABC Population Monte Carlo (ABC-PMC) sampling method to solve this problem [36]. Compared with the ABC-PMC sampling, which uses an adaptive Gaussian transition kernel, the ABC Sequence Monte Carlo (ABC-SMC) [37] sampling method has no restrictions on the transition kernel. In this study, the ABC Non-parametric Population Monte Carlo (ABC-NPMC) [38] sampling method based on the ABC-PMC with adaptive tolerance is used, which can make sampling achieve high efficiency.

However, with the increas of complexity and computational cost in product design, the popular uncertainty analysis algorithms seem to be difficult to handle such problems. Sheet metal forming is a such kind of problem and has been widely used in automobile, aviation, household electronics and other fields. For sheet metal forming design, it is a complex problem which is difficult to be handled due to its expensive computatial cost. In the past decades, the optimization and inverse methods have been widely used [39-44]. Commonly, most of these methods obtain a point estimation of design parameters. However, considering the uncertainty, the fluctuation of design parameters might lead to the diversity of forming quality, and the parameters corresponding to the required forming quality should be a range, rather than a value. Moreover, once the geometric shape of the forming part is determined, the main uncertainties affecting the forming quality are the material parameters of blank and the process parameters of forming process.With the increase of design varaibles, the computationl cost of should be increased significantly due to the curse of dimensionaly. Therefore, the VAE which can be applied dimension reduction in this study. Additonaly, It should be noted that the common forming defects in the sheet metal forming are wrinkling and cracking. In the existing research, the determination of parameters [45, 46] are mostly based on the forming criteria, such as thickness variation [47] and Forming Limit Diagram (FLD) [48, 49] which is difficult to fully reflect the forming quality. For example, these existing forming criteria focus on non-working areas that have little impact on formability, which is an interference to the determination of parameters. Furthermore, great differences in formability mapping to the objective function may be similar. Therefore, a more comprehensive evaluation criterion of sheet metal forming quality is needed. In the current work, the image (FLD) is processed directly as an objective function which can completely restract the characteristics of formed blank and reflect the forming quality.

In this study, the VAE is extended into the ABC in order to compromise the information loss and computational cost of the summary statistics. The relationship between design parameters and feature latent variables in the VAE is constructed by the LSSVR surrogate model. The design parameters will be identified by the VAE-based ABC method. For the proposed method, the posterior distribution of design parameters is obtained by ABC-NPMC sampling method. Taking sheet metal forming as an example, the material parameters and process parameters are identified by the proposed method. The remainder of this paper is organized as follows. In Section 2, Variational Auto-Encoder is introduced. In Section 3, the VAE-based ABC method is proposed. In Section 4, the design parameters of the sheet metal forming are identified by the proposed method. In the final Section, the conclusions are given.

## 2 Variational Auto-Encoder

Figure 1 shows the network structure of the VAE [50]. For a given dataset $\mathbf{D}=\{\mathbf{D}_i\}_{i=1}^{n}$, the distribution $P(\mathbf{D})$ of $\mathbf{D}$ can be obtained by introducing the distribution $P(\mathbf{Z})$ of the latent variable $\mathbf{Z}=\{\mathbf{Z}_i\}_{i=1}^{m}$. The equation is as follows:

$$P(\mathbf{D})=\sum_{\mathbf{Z}}P(\mathbf{D}\mid\mathbf{Z})P(\mathbf{Z}) \tag{1}$$

Here the reparameterization trick is used to determine $\mathbf{Z}$.

$$\mathbf{Z}=f(\mathbf{e},\mathbf{u})=\boldsymbol{\sigma}\odot\mathbf{e}+\mathbf{u} \tag{2}$$

where $\mathbf{e}$ is an independent auxiliary vector $\mathbf{e}\sim N(\mathbf{0},\mathbf{I})$. From the perspective of coding theory, the $P(\mathbf{D}\mid\mathbf{Z})$ acts as the decoder. Similarly, an arbitrary distribution $q(\mathbf{Z}\mid\mathbf{D})$ instead of true posterior $P(\mathbf{Z}\mid\mathbf{D})$ is introduced as the encoder. Eq. (3) and Eq. (4) give the encoding and decoding functions in the encoder and decoder.

$$\mathbf{Z}=f_{\vartheta}(\mathbf{D}) \tag{3}$$

$$\mathbf{Y}=g_{\gamma}(\mathbf{Z})\approx\mathbf{D} \tag{4}$$

where $\vartheta$ and $\gamma$ are parameter vectors of the encoder and decoder, respectively. Formally, as shown in Eq. (5), the KL divergence is used to measure the similarity between the approximate posterior $q(\mathbf{Z}\mid\mathbf{D})$ and the true posterior $P(\mathbf{Z}\mid\mathbf{D})$.

$$KL(q(\mathbf{Z}\mid\mathbf{D})\,||\,P(\mathbf{Z}\mid\mathbf{D}))=\log P(\mathbf{D})-L_b \tag{5}$$

where

$$L_b=-KL(q(\mathbf{Z}\mid\mathbf{D})\,||\,P(\mathbf{Z}))+E_{q(\mathbf{Z}|\mathbf{D})}[\log P(\mathbf{D}\mid\mathbf{Z})] \tag{6}$$

is the Evidence Lower Bound (ELBO) of the $\log P(\mathbf{D})$. For the given dataset $\mathbf{D}$, $P(\mathbf{D})$ is a constant and minimizing $KL(q(\mathbf{Z}\mid\mathbf{D})\,||\,P(\mathbf{Z}\mid\mathbf{D}))$ is equivalent to $L_b$ maximization. If $P(\mathbf{Z})$ is a standard Gaussian distribution $N(\mathbf{0},\mathbf{I})$ and $q(\mathbf{Z}\mid\mathbf{D})$ is a Gaussian distribution $N(\mathbf{u},\boldsymbol{\sigma}^2)$, the first term in $L_b$ can be calculated as:

$$KL(q(\mathbf{Z}\,|\,\mathbf{D})\,\|\,P(\mathbf{Z})) = \frac{1}{2}\sum_{i=1}^{m}(\sigma_i^2 - \log\sigma_i^2 + u_i^2 - 1) \tag{7}$$

For the second term in $L_b$, it is difficult to obtain an analytical expression, but it is actually equivalent to minimizing the difference between input $\mathbf{D}$ and output $\mathbf{Y} = \{\mathbf{Y}_i\}_{i=1}^{n}$. As shown in Eq. (8), the square error function is used here to evaluate this difference.

$$L(\mathbf{D},\mathbf{Y}) = \mathrm{L}(\mathbf{D}, \mathrm{g}_\gamma(\mathbf{Z})) = \|\mathbf{D}-\mathbf{Y}\|^2 \tag{8}$$

Based on Eq. (7) and Eq. (8), when the network has been completely trained, the optimized objective function is:

$$loss = KL(q(\mathbf{Z}\,|\,\mathbf{D})\,\|\,P(\mathbf{Z})) + L(\mathbf{D},\mathbf{Y}) = \frac{1}{2}\sum_{i=1}^{m}(\sigma_i^2 - \log\sigma_i^2 + u_i^2 - 1) + \|\mathbf{D}-\mathbf{Y}\|^2 \tag{9}$$

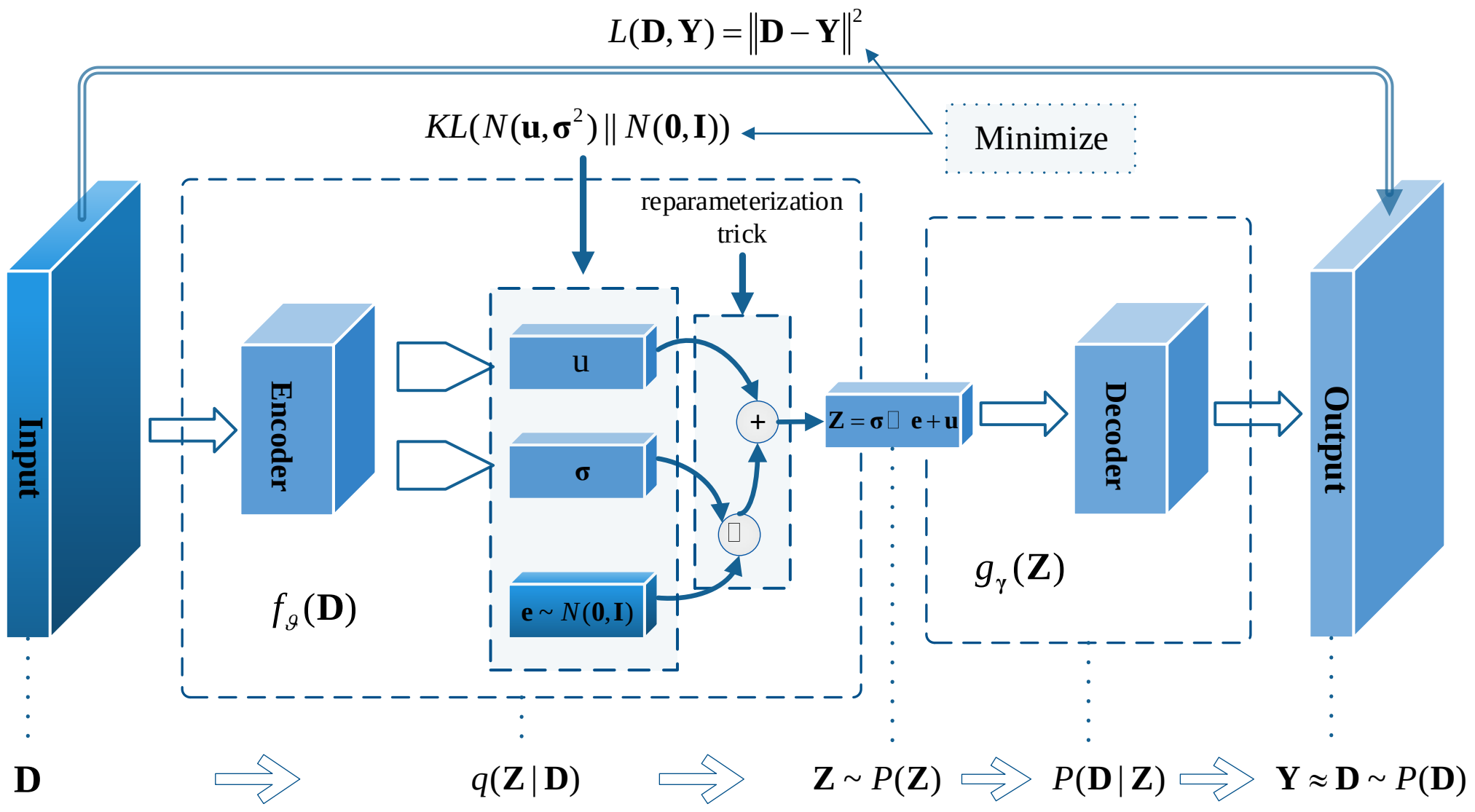

Fig. 1 The VAE network structure

When $P(\mathbf{D})$ is obtained, $\mathbf{Z}_i$ can be sampled arbitrarily in $P(\mathbf{Z})$, which is equivalent to sampling $\mathbf{D}_i$ in $P(\mathbf{D})$. Thus, the latent variable $\mathbf{Z}_i$ is crucial because it is another expression of $\mathbf{D}_i$. If $m<n$, it's a dimension reduction process, the network attempts to use $\mathbf{Z}$ to describe $\mathbf{D}$ on a smaller scale while minimizing information loss.

# 3 VAE-based ABC method

## 3.1 Approximate Bayesian Computation

Bayesian framework is widely used in inverse problems because it can provide more parameter information and its flexibility. In the Bayesian framework, the posterior distribution of parameters can be obtained by regularizing the ill-posed inverse problem with prior information. Its equation is as follows:

$$P(\boldsymbol{\theta} \mid \mathbf{D}) = P(\mathbf{D} \mid \boldsymbol{\theta}) P(\boldsymbol{\theta}) / P(\mathbf{D}) \tag{10}$$

where $\boldsymbol{\theta}$ is the design parameter vector, $P(\boldsymbol{\theta})$ is the prior probability of $\boldsymbol{\theta}$, $\mathbf{D}$ is the known data, $P(\mathbf{D})$ is marginal likelihood, $P(\mathbf{D} \mid \boldsymbol{\theta})$ is the likelihood function of $\boldsymbol{\theta}$, and $P(\boldsymbol{\theta} \mid \mathbf{D})$ is the posterior probability of $\boldsymbol{\theta}$. Normally, $P(\mathbf{D})$ can be normalized to a constant, so the Eq. (10) can be written as:

$$P(\boldsymbol{\theta} \mid \mathbf{D}) \propto P(\mathbf{D} \mid \boldsymbol{\theta}) P(\boldsymbol{\theta}) \tag{11}$$

For some complex models, it is tricky to obtain an analytical expression of the likelihood functions. To solve this problem, Pritchard et al. proposed an Approximate Bayesian Computation (ABC) instead of the Bayesian framework [19]. The main idea of ABC is to generate simulated data $\hat{\mathbf{D}}$ by some parameter vectors $\boldsymbol{\theta}^* = \boldsymbol{\theta}$. If the preset distance $\rho(\mathbf{D}, \hat{\mathbf{D}})$ between observed data $\mathbf{D}$ and simulated data $\hat{\mathbf{D}}$ satisfies a tolerance $\varepsilon$, it indicates that simulated data $\hat{\mathbf{D}}$ is accepted, and $\boldsymbol{\theta}^*$ is retained as sample data. The equation is as follows:

$$\rho(\mathbf{D}, \hat{\mathbf{D}}) \le \varepsilon \tag{12}$$

Here, the Euclidean distance $\|\square\|$ is used to compare the distance between $\mathbf{D}$ and $\hat{\mathbf{D}}$. If the $\varepsilon$ is very small, the posterior $P(\boldsymbol{\theta} \mid \mathbf{D})$ can be approximated to $P(\boldsymbol{\theta}^* \mid \rho(\mathbf{D}, \hat{\mathbf{D}}) \le \varepsilon)$.

$$P(\boldsymbol{\theta}\,|\,\mathbf{D}) \approx P(\boldsymbol{\theta}^*\,|\,\rho(\mathbf{D},\hat{\mathbf{D}}) \le \varepsilon) = P(\boldsymbol{\theta}^*\,|\,\left\|\mathbf{D},\hat{\mathbf{D}}\right\| \le \varepsilon) \tag{13}$$

Generally, with the data dimension increases, it becomes more difficult to generate a simulated data $\hat{\mathbf{D}}$ that has a small distance from the observed data **D**. Formally, the low-dimensional summary statistic $S(\mathbf{D})$ can be used instead of the high-dimensional data **D**. So, the Eq. (13) changes to the following form:

$$P(\boldsymbol{\theta}\,|\,\mathbf{D}) \approx P(\boldsymbol{\theta}^*\,|\,\left\|\mathbf{D},\hat{\mathbf{D}}\right\| \le \varepsilon) = P(\boldsymbol{\theta}^*\,|\,\left\|S(\mathbf{D}),S(\hat{\mathbf{D}})\right\| \le \varepsilon) \tag{14}$$

If the summary statistic $S(\mathbf{D})$ is sufficient for the $\boldsymbol{\theta}$, $S(\mathbf{D})$ will capture all the information about $\boldsymbol{\theta}$ in the data **D**, then the calculation efficiency will be improved in this way without any error. Thus, the selection of summary statistics is the key to ABC implementation. If the scale of the summary statistics is too small, it may contain only a small part of information for $\boldsymbol{\theta}$, and the result of the calculation will be inaccurate. Conversely, if the scale of the summary statistics is too large, it will cost expensive calculations even though with a lot of parameter information.

### 3.2 VAE-based ABC framework

To make an effective trade-off between information loss and computational cost, the VAE is extended to ABC. The VAE has shown advantages in extracting features and reconstructing samples [28-31, 51, 52]. With the VAE, the high-dimensional observations can be projected into a low-dimensional latent space with minimal information loss. Obviously, the latent variable **Z** in VAE can effectively meet the two requirements of summary statistics. Therefore, it is appropriate to extend the VAE into ABC. The latent variable **Z** can be determined by Eq. (3). Meanwhile, the high-dimensional data can be explicitly represented by low-dimensional **Z** according to Eq. (4). Therefore, the Eq. (14) is rewritten as follows:

$$P(\boldsymbol{\theta}\,|\,\mathbf{D}) \approx P(\boldsymbol{\theta}^*\,|\,\left\|\mathbf{D},\hat{\mathbf{D}}\right\| \le \varepsilon) = P(\boldsymbol{\theta}^*\,|\,\left\|\mathrm{g}_\gamma(\mathbf{Z}),\mathrm{g}_\gamma(\hat{\mathbf{Z}})\right\| \le \varepsilon) \tag{15}$$

For a well trained VAE model, its decoding function $\mathrm{g}_\gamma(\mathbf{Z})$ is deterministic and the observed data **D** is only linearly related to **Z**. So, Eq. (15) can be rewritten as:

$$P(\boldsymbol{\theta} \mid \mathbf{D}) \approx P(\boldsymbol{\theta}^* \mid \left\| \mathbf{Z}, \hat{\mathbf{Z}} \right\| \leq \varepsilon) \tag{16}$$

where **Z** and $\hat{\mathbf{Z}}$ present the observed coefficient vector and the simulated coefficient vector of the ABC, respectively. In order to implement the ABC, the simulated coefficient vector under a given parameter vector needs to be obtained by Finite Element (FE) simulation which is very time-consuming. For the ABC, a large number of samples are required, which will result in prohibitive computation. Therefore, the nonlinear relationship between parameter vectors and latent variables is constructed by LSSVR surrogate model instead of the time-consuming FE simulation. The LSSVR surrogate model is a common tool for establishing complex input-output connections, which can construct more stable and accurate nonlinear relationships with a small number of samples [53]. The derivation of the LSSVR model is detailed in Appendix A. Here, the LSSVR model constructed based on latent variable is called LV-LSSVR, which is shown in Eq.(17).

$$\hat{\mathbf{Z}} = \sum_{i=1}^{m} \alpha_i K(\boldsymbol{\theta}, \boldsymbol{\theta}_i) + b \tag{17}$$

where $\alpha_i$ is the Lagrange multiplier, $b$ is the model bias, $K(\boldsymbol{\theta}, \boldsymbol{\theta}_i)$ is the kernel function matrix. With the well-constructed LV-LSSVR model, the simulated coefficient vectors under given parameter vectors can be obtained. Then, the design parameter values are determined by comparing the distance between the observed coefficient vectors and the simulated coefficient vectors according to Eq. (16). This method bypasses the estimation of the intractable likelihood function and makes the calculation more efficient by the LV-LSSVR model.

In order to obtain the posteriori probability distribution of the forming parameters, the sampling method is critical. Several sampling strategies, such as ABC-rejection, ABC-MCMC, ABC-PRC, ABC-PMC and ABC-SMC have been widely used. However, if an inappropriate sampling method is chosen, the sampling might not be easier to converge. Among them, the ABC-PMC sampling strategy is a competitive one and optimizes sampling acceptance probability with an adaptive Gaussian transition kernel $q(*|\boldsymbol{\theta}_i^t, \sigma_{t-1})$ [20]. Assuming that the number of iterations is $T$ and each particle pool contains $N$ particles. In the $t$-th ($1<t<T$) iteration, the sample $\boldsymbol{\theta}_i^t$ ($1<i<N$) is obtained with the weight $\mathbf{w}_i^{t-1}$ by the $q(\boldsymbol{\theta}_i^{t-1}|\boldsymbol{\theta}_i^t, \sigma_{t-1})$ of $t$-1th iteration to perturb $\boldsymbol{\theta}_i^{t-1}$. Meanwhile, the simulated coefficient vector $\hat{\mathbf{Z}}_i$ is generated with the $\boldsymbol{\theta}_i^t$. Comparing $\mathbf{Z}$ and $\hat{\mathbf{Z}}_i$, if $\rho(\mathbf{Z}, \hat{\mathbf{Z}}_i) \le \varepsilon$ is satisfied, the $\boldsymbol{\theta}_i^t$ is retained. The weight $\mathbf{w}_i^t$ and variance $\sigma_t^2$ are given as:

$$\mathbf{w}_i^t = \frac{P(\boldsymbol{\theta}_i^t)}{\sum_{j=1}^{N} \mathbf{w}_j^{t-1} q(\boldsymbol{\theta}_j^{t-1}|\boldsymbol{\theta}_i^t, \sigma_{t-1})} \tag{18}$$

$$\sigma_t^2 = 2 \cdot \frac{1}{N} \sum_{i=1}^{N} (\boldsymbol{\theta}_i^t - \sum_{j=1}^{N} \boldsymbol{\theta}_j^t / N)^2 = 2 \cdot Var(\boldsymbol{\theta}_{1:N}^t) \tag{19}$$

In this study, the ABC-NPMC sampling method based on the ABC-PMC is used. In the ABC-PMC, all information is known except the decreasing sequence of tolerance $\varepsilon$. Unlike the ABC-PMC, the ABC-NPMC sampling method can adaptively determine the $\varepsilon$, which is helpful for sampling. The details can be found in Ref.[38].

The flowchart of the VAE-based ABC method is shown in Fig. 2. The details of calculation process are presented as follows:

a. Generate the training samples $\boldsymbol{\theta}_{\text{tra}}$ and test samples $\boldsymbol{\theta}_{\text{test}}$ in parameter space by Latin Hypercube Design (LHD) as shown in Fig. 2 (a).
b. Perform FE evaluations for both training and test samples to obtain training and test images respectively.
c. Reconstruct an objective image by the pixels of the training images.
d. As shown in Fig. 2 (b), the VAE network is trained with the training images and corresponding objective one. The latent variables **Zs** and **Zo** corresponding to the training images and the objective image should be obtained after the training.

e. Construct the LV-LSSVR model with the training samples $\boldsymbol{\theta}_{\mathrm{tra}}$ and latent variables $\mathbf{Zs}$ as shown in Fig. 2 (c).

f. Obtain responses $\mathbf{Zt}$ of the test samples by the constructed LV-LSSVR model as shown in Fig. 2 (d).

g. Decode the corresponding pseudo images with responses $\mathbf{Zt}$ by the trained decoder.

h. Evaluate the accuracy of LV-LSSVR model with decoded pseudo images and test images from FE evaluations. If accuracy is met, procedure goes to Step *i*th. Otherwise, add 200 new training samples by the LHD and goes back to step *a*.

i. The parameter vector in parameter space should be given, and obtain the simulated coefficient vector $\hat{\mathbf{Z}}$ with the well-constructed LV-LSSVR model.

j. Compare the observed coefficient vector $\mathbf{Zo}$ and simulated coefficient vector $\hat{\mathbf{Z}}$ to complete the VAE-based ABC calculation with the ABC-NPMC sampling as shown in Fig. 2 (e).

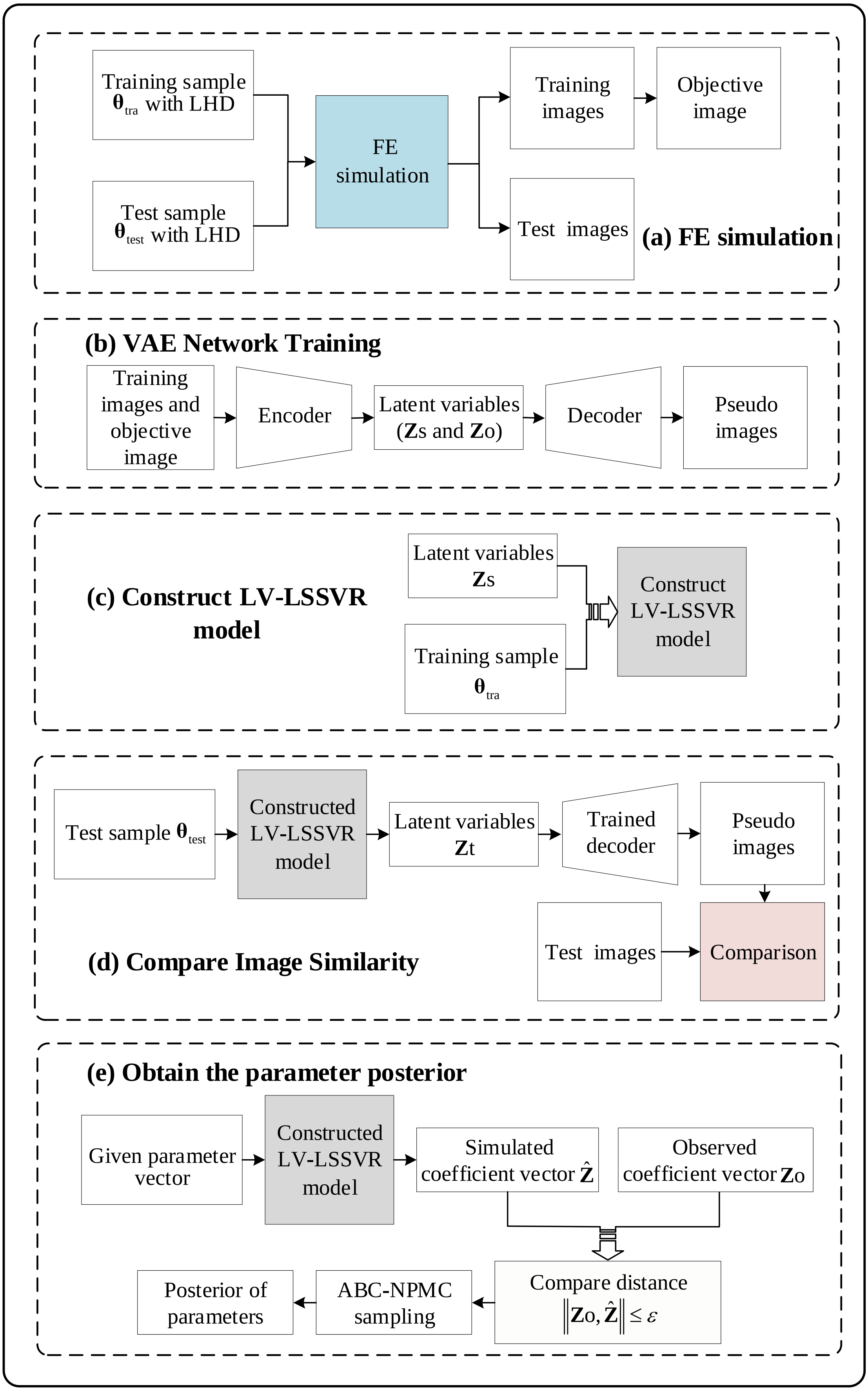

Fig. 2 The flowchart of the VAE-based ABC method

# 4 Identification of sheet metal forming parameters

Due to the increasing complexity of sheet metal forming, it is usually difficult to obtain a forming part without cracking and wrinkling effectively in practical engineering. Generally, a lot of simulations and experiments are required to find an approximately appropriate forming parameters which will greatly increase the design cycle of the product. Therefore, the parameter identification is needed. However, the parameter identification of forming process is a high computational cost problem. Moreover, it is difficult to set an effective objective function to reflect the forming quality. Therefore, the VAE-based ABC method that can handle such problems is proposed.

In the existing methods, several criteria are used as objective function to reflect the forming quality [42], such as thickness reduction, FLD and springback. The thickness reduction is allowed as a formability criterion because a crack is always preceded by a high thinning and wrinkling is always preceded by a high thickening in practice [54]. The thinning rate between the initial and final states is given by Barlet [47], et al, as

$$f_h = \left( \sum_{e=1}^{N} f_h(e) \right)^{1/p} \tag{20}$$

where

$$f_h(e) = \left( \frac{h_e - h_0}{h_0} \right)^p \tag{21}$$

where $h_0$ and $h_e$ are the initial and final thicknesses, respectively; $N$ is the number of elements in a blank FE model; The coefficient $p = 2, 4, 6 \cdots$ is introduced to emphasize the extremes of the objective function. Compared with thickness reduction, the FLD [48, 55] provides a graphical description of material failure test and is widely used. Breitkopf et al. defined two forming limit curves (FLCs) [49] in the principal plan of logarithmic strain as shown in Eq.(22).

$$\begin{aligned} \varepsilon_1 &= \varphi_s(\varepsilon_2) \\ \varepsilon_1 &= \varphi_w(\varepsilon_2) \end{aligned} \tag{22}$$

where $\varphi_s$ and $\varphi_w$ are used to control crack and wrinkle, respectively; $\varepsilon_1$ and $\varepsilon_2$ are major strain and minor strain, respectively. Therefore, a safety FLC is defined as follows:

$$\begin{aligned} \vartheta_s(\varepsilon_2) &= \varphi_s(\varepsilon_2) - s \\ \vartheta_w(\varepsilon_2) &= \varphi_w(\varepsilon_2) - s \end{aligned} \tag{23}$$

where $s$ is the allowable safe distance. The final equation is as follows:

$$f_\varepsilon = \left( \sum_{e=1}^{N} f_\varepsilon(e) \right)^{1/p} \tag{24}$$

where

$$\begin{cases} f_\varepsilon(e) = \left( \varepsilon_1^e - \vartheta_s(\varepsilon_2^e) \right)^p & \text{for } \varepsilon_1^e > \vartheta_s(\varepsilon_2^e), \\ f_\varepsilon(e) = \left( \vartheta_w(\varepsilon_2^e) - \varepsilon_1^e \right)^p & \text{for } \varepsilon_1^e < \vartheta_w(\varepsilon_2^e), \\ f_\varepsilon(e) = 0 & \text{otherwise} \end{cases} \tag{25}$$

The springback [56, 57] is commonly measured by the distance between nodes on a drawn part before and after springback. The springback criterion can be used for regular shapes, for complex geometric shapes, how to establish a reasonable criterion remains to be further studied. Actually, due to the complexity of sheet metal forming design, the existing criteria are difficult to completely reflect the forming quality. Therefore, the image (FLD) is processed directly as objective function in this study, which can completely restract the characteristics of formed blank and reflect the forming quality. In this section, the proposed method is used to identify the sheet metal forming parameters and the details are presented as follows.

### 4.1 FE model

In this case, LSDYNA is used for simulation. The forming part is an engine inner hood and the CAD model as shown in Fig. 3. The FE model is composed of blank and tools (die, punch and binder). The blank is modeled by 8,066 quadrilateral elements. The blank thickness is 0.8*mm* and the material is DC04_0.80 *mm* (36). The tools are modeled by 81,740 quadrilateral elements and 28,728 triangular elements. Each simulation time is about 90 minutes when using the personal computer (Inter(R) Core (TM) i5-7400 CPU, 3.00GHz, RAM 12GB). Obviously, the FE simulation is time-consuming.

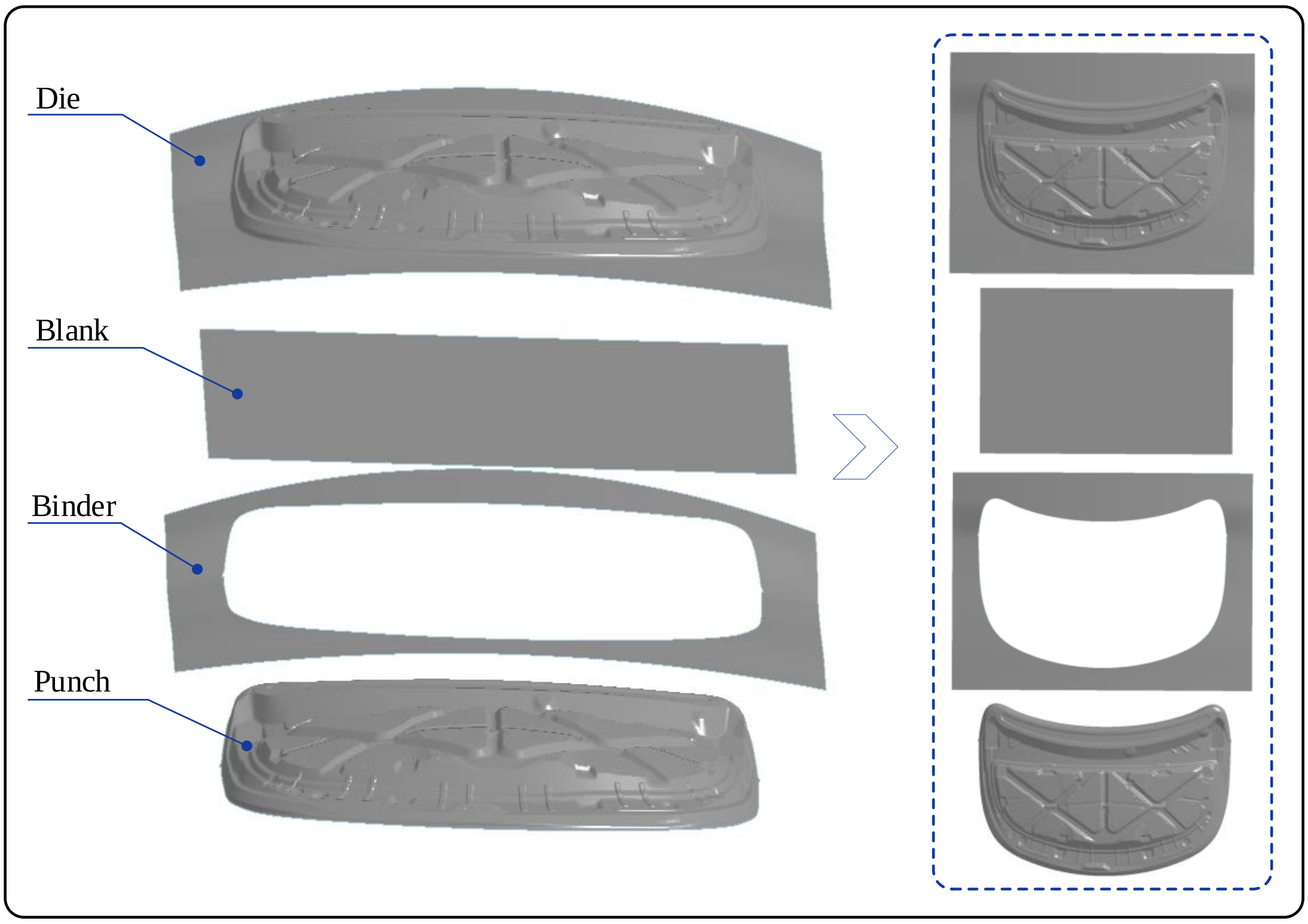

Fig. 3 The CAD model of engine inner hood

The final formed FLD is shown in Fig. 4. Its FLD can be parttioned as 6 regions involving crack, risk of crack, safe, insufficient stretch, wrinkle tendency and wrinkles. For each region, the corresponding color is displayed. The crack, wrinkles and safe areas are mainly considered. For sheet metal forming, it is necessary to maximize the number of elements in the safe area and the elements without crack and wrinkles. Considering the uncertainty, the parameters corresponding to promising forming quality should be within a certain range. Therefore, it is necessary to identify the material parameters and process parameters, and find the optimum parameter range for the forming result. In the current work, a total of 12 material parameters and process parameters to be identified are listed in Table 1.

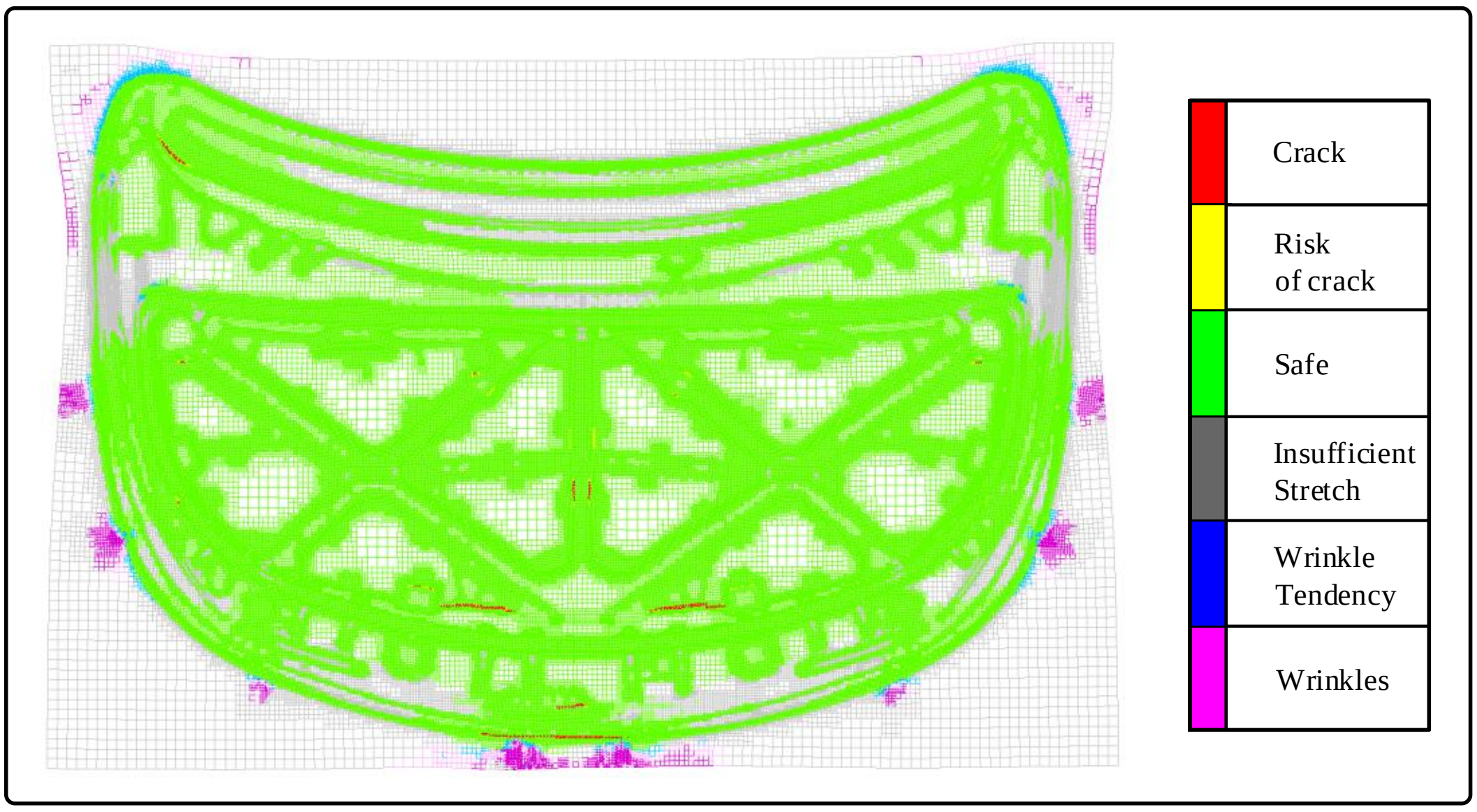

Fig. 4 The results of FE simulation

**Table 1** The material parameters and process parameters

| Material parameters | | Process parameters | |
|---|---|---|---|
| Young's modulus (*Gpa*) | $E$ | Friction coefficient of punch/die | $f_1$ |
| Poisson's ratio | $u$ | Friction coefficient of binder | $f_2$ |
| Strength coefficient | $K$ | Drawing speed (*mm*/*s*) | $v$ |
| Hardening exponent | $N$ | Blank holder force (*Ton*) | BHF |
| Lankford parameter | $R_{00}$ | Drawbead resistance (*N*) | $F$ |
| Lankford parameter | $R_{45}$ | | |
| Lankford parameter | $R_{90}$ | | |

## 4.2 Image preprocessing

As for sheet metal forming parts, the forming quality can be determined directly by the FLD. However, as mentioned before, the FLD not only describes the working region, but also expresses the forming performance of the non-working region with little influence on the working region. Therefore, the image processing technique based on OpenCV is applied to segment the working region and non-working region, and the working region is studied in current work. Then, the objective image is reconstructed from the existing ones, which is beneficial to the subsequent parameter identification. In this work, 800 parameter vectors are sampled by the LHD as training samples, and the FLDs are obtained by FE simulation with these samples. The non-working region of blank after forming is not considered. Then, the mask based on the punch border as shown in Fig.5 is employed to obtain the expected images for training.

**Table 2** The image preprocess procedure

```
Obtian a mask:
Input: image of punch;
Output: a mask for segment working region;
-  Convert image from RGB to HSV;
-  Color_low= Minimum h, s, v value of punch color area;
-  For each pixel:
       If   h, s, v>Color_low
           h, s, v = 0, 0, 0;
       Else
           h, s, v = 0, 0, 255;
Process the FLD:
Input: mask and FLDs;
Output: the FLDs with only working region;
-  For image in FLDs:
       a=RGB value of image; b=RGB value of mask;
       Processed_FLD=a+b;
Reconstruct an objective image:
Input: processed FLDs;
Output: an all-green objective image;
-  Objective_ image= one of the processed FLDs;
-  Convert Objective_ image from RGB to HSV;
-  Green _low= Green minimum h, s, v value;
-  Green _up= Green maximum h, s, v value;
-  For image in processed FLDs:
       Convert image from RGB to HSV;
       For each pixel:
           a= Objective_ image h, s, v value;
           b= Objective_ image h, s, v value;
           If Green _low < a <Green _up
                Nothing;
           Else
                If Green _low < b <Green _up
                    a=b;
-  Convert Objective_ image from HSV to RGB;
```

Actually, a perfect forming performance which the FLD is all green in working region is usually dose not require FE evluations. Therefore, considering the different values of green pixels in different positions in the image, a new objective image can be generated by pixel replacement based on the given evluated 800 training images. Firstly, an image is selected from the existing image set. Then, traversing the entire image set, the non-green pixels in the selected image are replaced by the green pixels in the same position on other images to generate the objective image instead of the FE evalution. The image processing process is shown in Fig. 5. The detailed process is shown in Table 2.

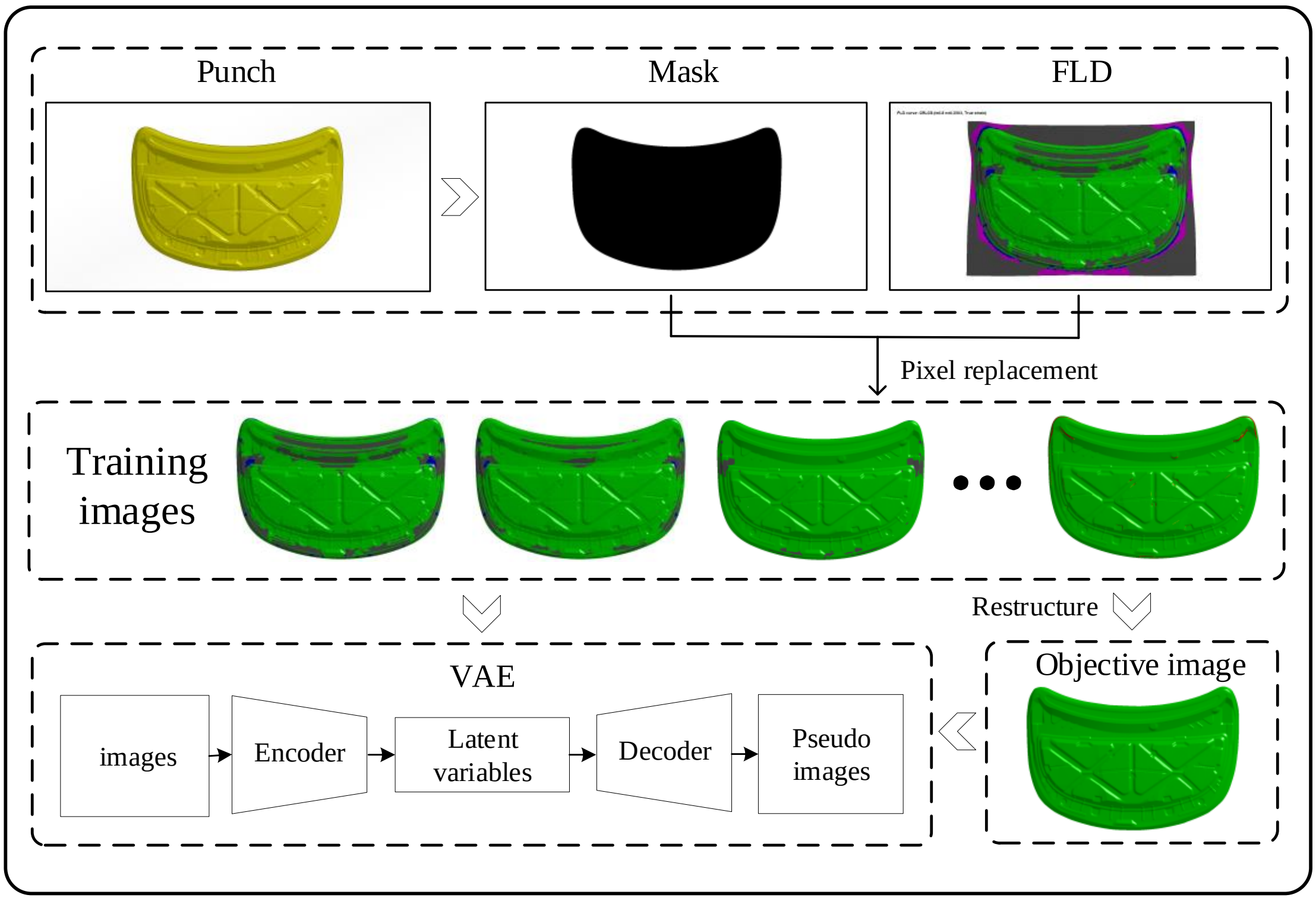

Fig. 5 The image processing for sheet forming problem

### 4.3 Identification of parameters

As shown in Fig. 6, the input of the encoder of VAE is the image to be trained, and the output is the one-dimensional mean $u$ and Standard Deviation (STD) $\sigma$. The decoder consists of 6 Batch Normalized-Convolution layers (BN-Conv) and one full connection layer. Batch normalization allows training to use higher learning rates without too much consideration of the initialization [58]. For a given layer of $d$ dimension input $\mathbf{X}=\{x_i\}_{i=1}^{d}$, each dimension is normalized by batch normalization as follows:

$$y_i = \gamma x_i' + \beta \tag{26}$$

where

$$x_i' = \frac{x_i - u_x}{\sigma_x} \tag{27}$$

subject to

$$\begin{aligned} u_x &= \frac{1}{d}\sum_{i=1}^{d} x_i \\ \sigma_x &= \frac{1}{d}\sum_{i=1}^{d} (x_i - u_x)^2 \end{aligned} \tag{28}$$

where $\gamma$ and $\beta$ are parameters to be learned; $u_x$ and $\sigma_x$ are the mean and STD of **X**. The Leaky Rectified Linear Unit (LReLU) function as shown in Eq. (29) is used as the activation function of the encoder.

$$LRelu = \max(x, \lambda x) \tag{29}$$

where $\lambda$ is a value between 0 and 1, which is set to 0.2 from reference [59]. The input and output sizes, filter shape, strides, and padding rule for each layer of the encoder are shown in Table 3.

**Table 3** The structure of encoder

| | **Input size** | **Output size** | **Filter shape** | **Strides** | **Padding rule** |
|---|---|---|---|---|---|
| BN-Conv 1 | 256x256x3 | 128x128x32 | (4,4,3,32) | [1,2,2,1] | SAME |
| BN-Conv 2 | 128x128x32 | 64x64x64 | (4,4,32,64) | [1,2,2,1] | SAME |
| BN-Conv 3 | 64x64x64 | 32x32x128 | (4,4,64,128) | [1,2,2,1] | SAME |
| BN-Conv 4 | 32x32x128 | 16x16x256 | (4,4,128,256) | [1,2,2,1] | SAME |
| BN-Conv 5 | 16x16x256 | 8x8x512 | (4,4,256,512) | [1,2,2,1] | SAME |
| BN-Conv 6 | 8x8x512 | 4x4x512 | (4,4,512,512) | [1,2,2,1] | SAME |
| Full connection 1 | 1x8192 | 1x1($u$ and $\sigma$) | | | |

The input of the decoder in VAE is one-dimensional feature latent variable ***Z*** obtained from Eq.(2), and the output is the pseudo image. Similarly, the decoder consists of a full connection layer and six transposed convolution layers. The Rectified Linear Unit (ReLU) function as shown in Eq.(30) is used as the activation function of the decoder.

The input and output sizes, filter shape, strides, and padding rule for each layer of the decoder are shown in Table 4.

$$Relu = \max(x, 0) \tag{30}$$

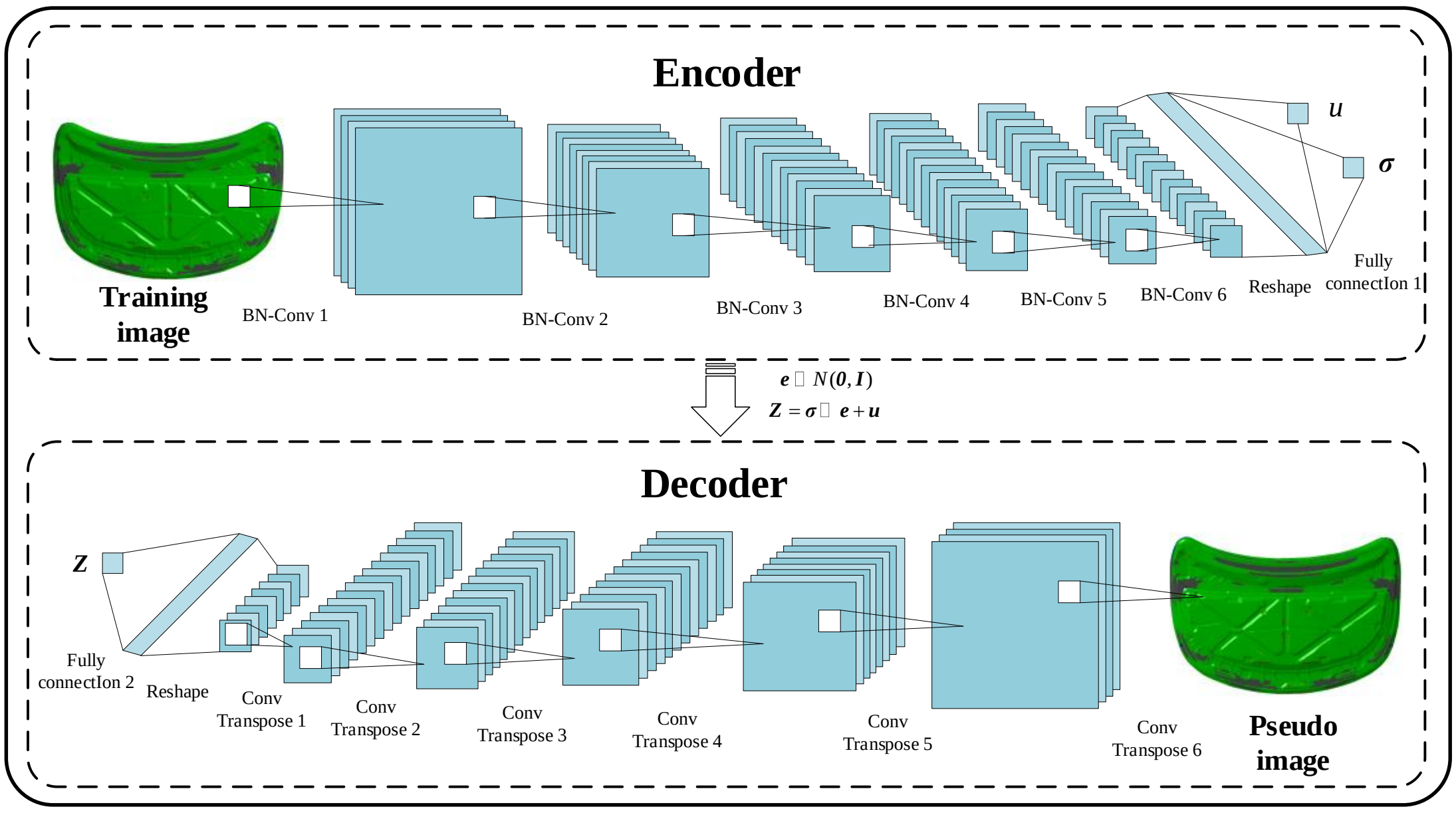

Fig. 6 The architecture of the VAE

**Table 4** The structure of decoder

| | **Input size** | **Output size** | **Filter shape** | **Strides** | **Padding rule** |
|---|---|---|---|---|---|
| Full connection 2 | 1x1(z) | 1x512 | | | |
| Conv transpose 1 | 4x4x32 | 8x8x512 | (4,4,512,32) | [1,2,2,1] | SAME |
| Conv transpose 2 | 8x8x512 | 16x16x256 | (4,4,256,512) | [1,2,2,1] | SAME |
| Conv transpose 3 | 16x16x256 | 32x32x128 | (4,4,128,256) | [1,2,2,1] | SAME |
| Conv transpose 4 | 32x32x128 | 64x64x64 | (4,4,64,128) | [1,2,2,1] | SAME |
| Conv transpose 5 | 64x64x64 | 128x128x32 | (4,4,32,64) | [1,2,2,1] | SAME |
| Conv transpose 6 | 128x128x32 | 256x256x3 | (4,4,3,32) | [1,2,2,1] | SAME |

The optimizer used in the VAE is the Adaptive Moment Estimation optimizer (Adam) [60], which is essentially the Root Mean Square Prop (RMSProp) [61] with momentum factor. The Adam integrates the advantages of Adaptive Gradient (AdaGrad) [62] and RMSProp, and has lower computing cost. Besides, it performs well in most non-convex optimization, large data sets and high-dimensional space [59]. In the current work, the VAE training process sets 150 epochs, and each epoch trains 801 images (800 training images and 1 objective image). The average loss value of each epoch is shown in Fig. 7. It can find that the images trained in the 1st epoch is blurred, the complete contour of the images can be learned in the 10th epoch, the images have been trained well in the 70th epoch, and the images are trained completely in the last epoch.

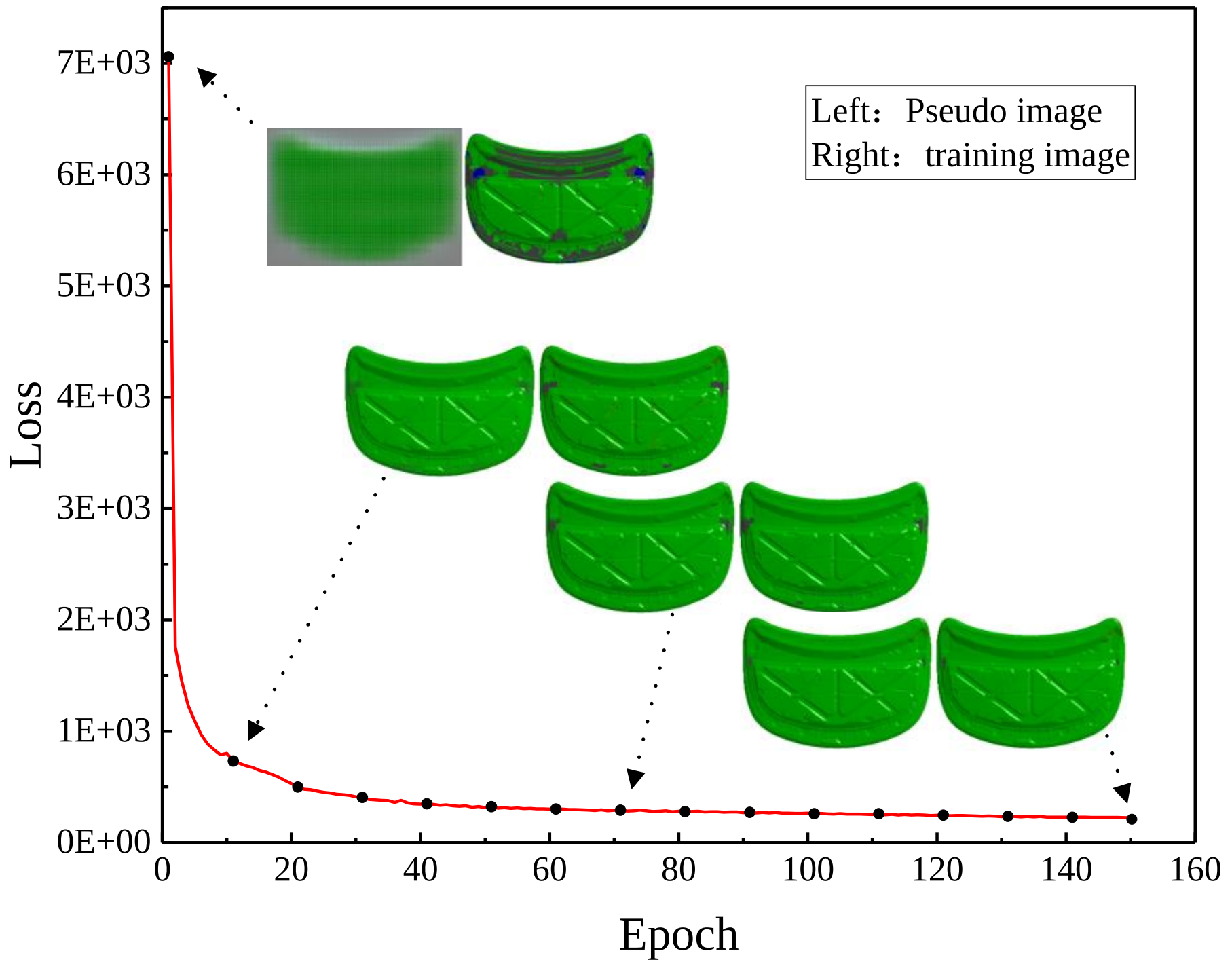

Fig. 7 The average loss value for each epoch

Figure 8 shows some training images, an objective image and their corresponding pseudo images (which are trained by the VAE instead of the FE evaluation) after training. It can be found that there is no significant difference between the pre-training and the trained pseudo images, indicating that most of the features of the image have been learned. After the training is completed, the feature latent variables corresponding to the training images and the objective image are $\mathbf{Zs}$ and $\mathbf{Zo}$, respectively. Meanwhile, each feature latent variable corresponds to a trained pseudo image. Furthermore, based on the characteristics of the VAE reconstructed samples, when a new latent variable $\hat{\mathbf{Z}}$ is given, a pseudo image similar to the real image is generated. Addtionaly, in order to further improve computational efficiency, the LV-LSSVR model is constructed by training samples $\boldsymbol{\theta}_{tra}$ and $\mathbf{Zs}$ instead of the time-consuming FE simulation. With an accurate LV-LSSVR model constructed, $\hat{\mathbf{Z}}$ can be predicted with the given parameter vector $\boldsymbol{\theta}$.

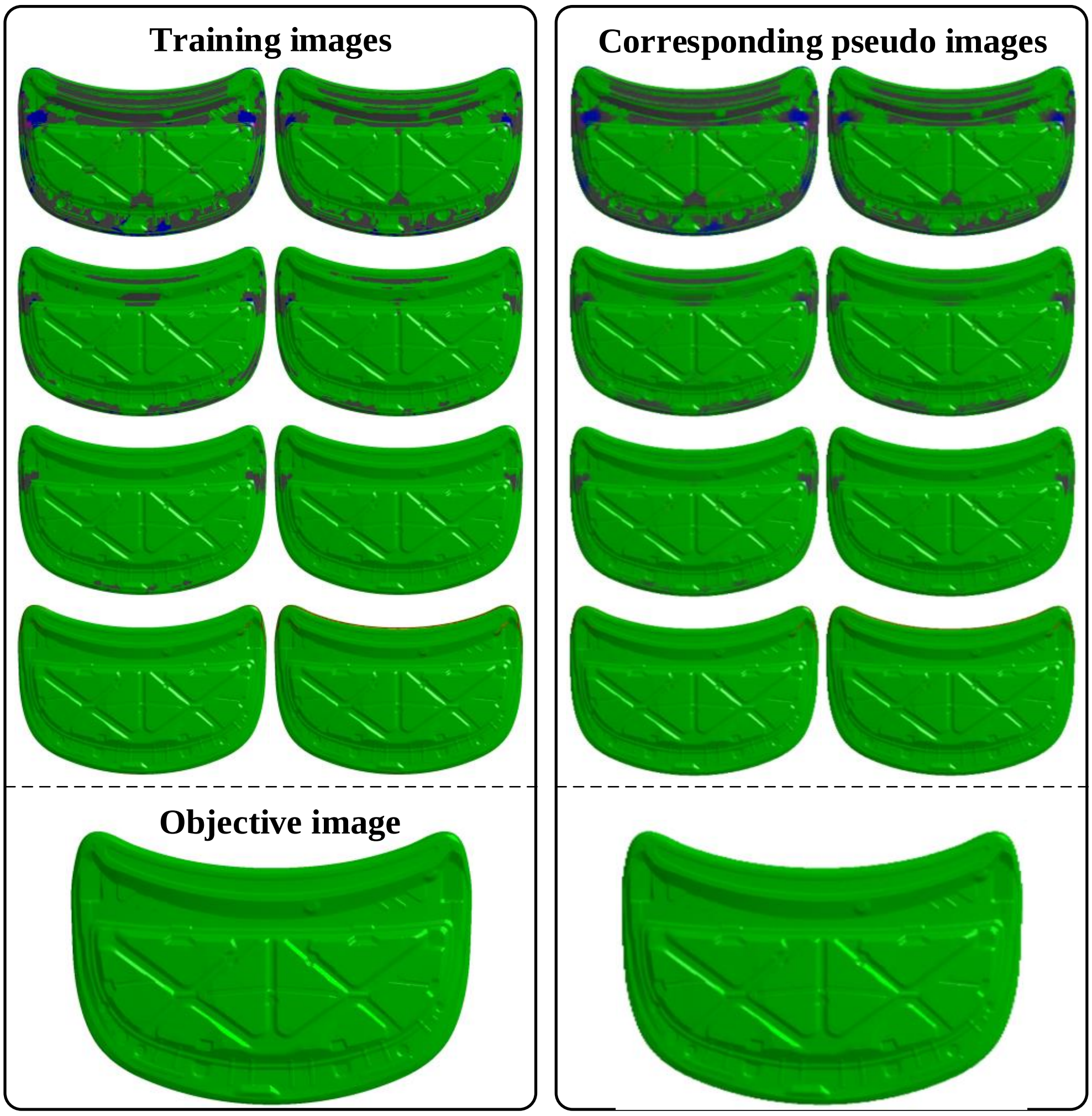

Fig. 8 The images and corresponding pseudo images

In order to test the accuracy of the model, 350 test samples $\boldsymbol{\theta}_{\text{test}}$ are sampled by the LHD in the design space. The test images are obtained by the FE simulations with these test samples. Meanwhile, the latent variables $\mathbf{Zt}$ corresponding to the test samples are obtained by the constructed LV-LSSVR model and $\mathbf{Zt}$ are decoded to the corresponding pseudo images by the decoder. Comparing the similarity between the pseudo and test images, the Structural Similarity (SSIM) [63] is used to compare image similarity. In the SSIM, the differences of the luminance, contrast, and structure of the image are considered comprehensively, so it is widely used [64]. For two images $\mathbf{x}$, $\mathbf{y}$,

their SSIM is:

$$\text{SSIM}(\mathbf{x},\mathbf{y})=[l(\mathbf{x},\mathbf{y})]^{\alpha}\cdot[c(\mathbf{x},\mathbf{y})]^{\beta}\cdot[s(\mathbf{x},\mathbf{y})]^{\gamma} \tag{31}$$

where $\alpha > 0, \beta > 0$ and $\gamma > 0$ are parameters used to adjust the relative importance of the components; The three components $l(\mathbf{x},\mathbf{y})$, $c(\mathbf{x},\mathbf{y})$ and $s(\mathbf{x},\mathbf{y})$ (Eq.(32))are luminance, contrast and structure comparisons respectively.

$$\begin{aligned} l(\mathbf{x},\mathbf{y}) &= \frac{2u_x u_y + C_1}{u_x^2 + u_y^2 + C_1} \\ c(\mathbf{x},\mathbf{y}) &= \frac{2\sigma_x \sigma_y + C_2}{\sigma_x^2 + \sigma_y^2 + C_2} \\ s(\mathbf{x},\mathbf{y}) &= \frac{\sigma_{xy} + C_3}{\sigma_x \sigma_y + C_3} \end{aligned} \tag{32}$$

where $u_x$ and $u_y$ are the mean of **x** and **y**, respectively; $\sigma_x$ and $\sigma_y$ are the STD of **x** and **y**, respectively; $\sigma_{xy}$ is the covariance of **x** and **y**; $C_1$, $C_2$ and $C_3$ are constant to avoid system errors caused by denominator being very close to zero. In engineering practice, generally set to $\alpha=\beta=\gamma=1$ and $C_3 = 0.5C_2$, and Eq. (31) is simplified to

$$\text{SSIM}(\mathbf{x},\mathbf{y}) = \frac{(2u_x u_y + C_1)(2\sigma_{xy} + C_2)}{(u_x^2 + u_y^2 + C_1)(\sigma_x^2 + \sigma_y^2 + C_2)} \tag{33}$$

The SSIM is a value between 0 and 1, and the larger the SSIM, the smaller the difference between the two images. Figure 9 shows the SSIM of 350 sets of test images and corresponding pseudo images, and the mean of these SSIMs is 0.9193. It can be found that all SSIM values are above 0.8, and most of them are above 0.9. It means that most of the test images and the corresponding pseudo images are not much different. Figure 10 shows the partial comparison results of the images. Therefore, the LV-LSSVR model can be used to predict latent variables with parameter vectors. Meanwhile, the latent variables can be used instead of images as summary statistics for the ABC. Sequentially, the latent variable **Zo** corresponding to the objective image is used as the observed coefficient vector of ABC, and the simulated coefficient vector $\hat{\mathbf{Z}}$ is generated by the LV-LSSVR model with the given parameter vector.

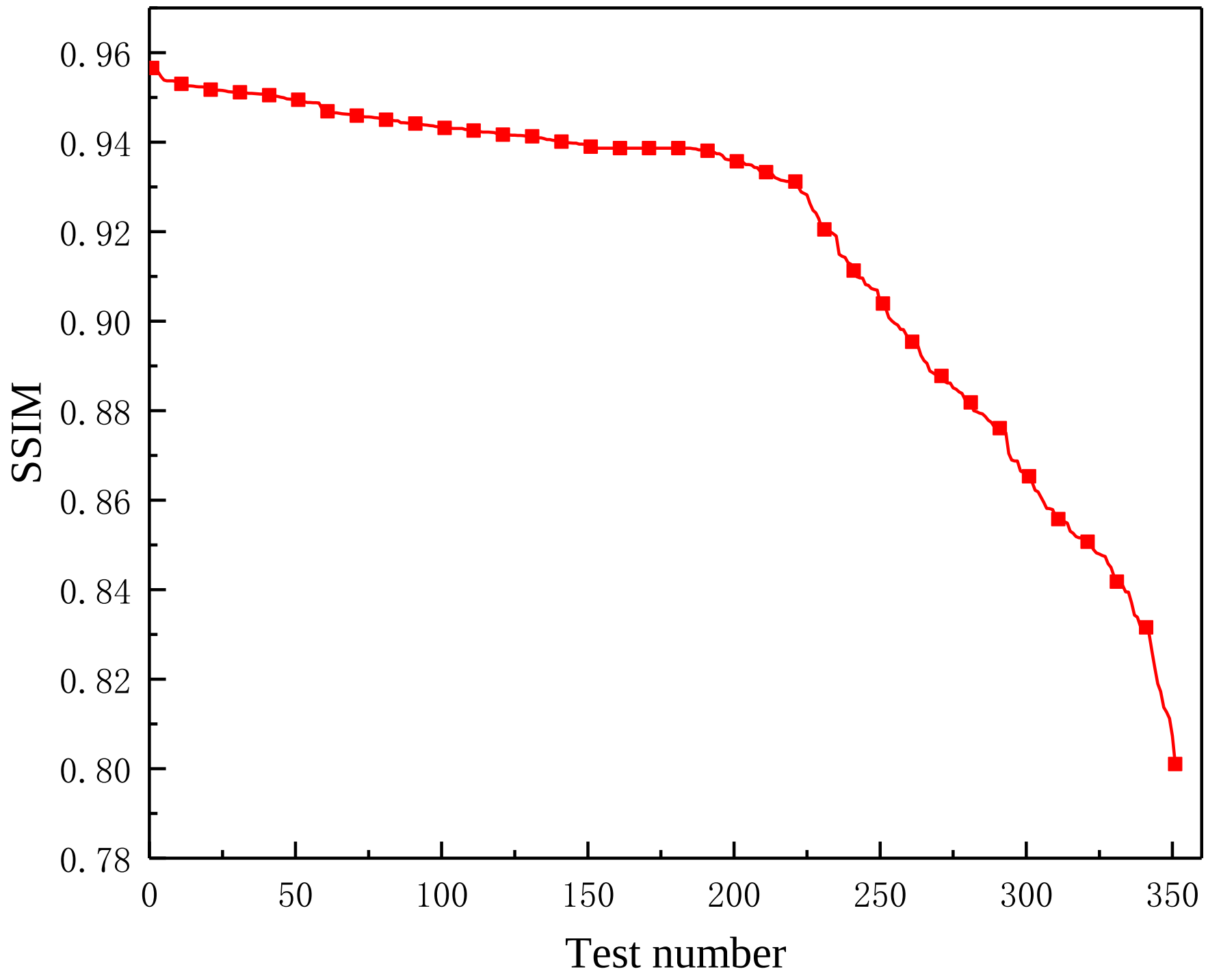

Fig. 9 The SSIM values of test images and corresponding pseudo images

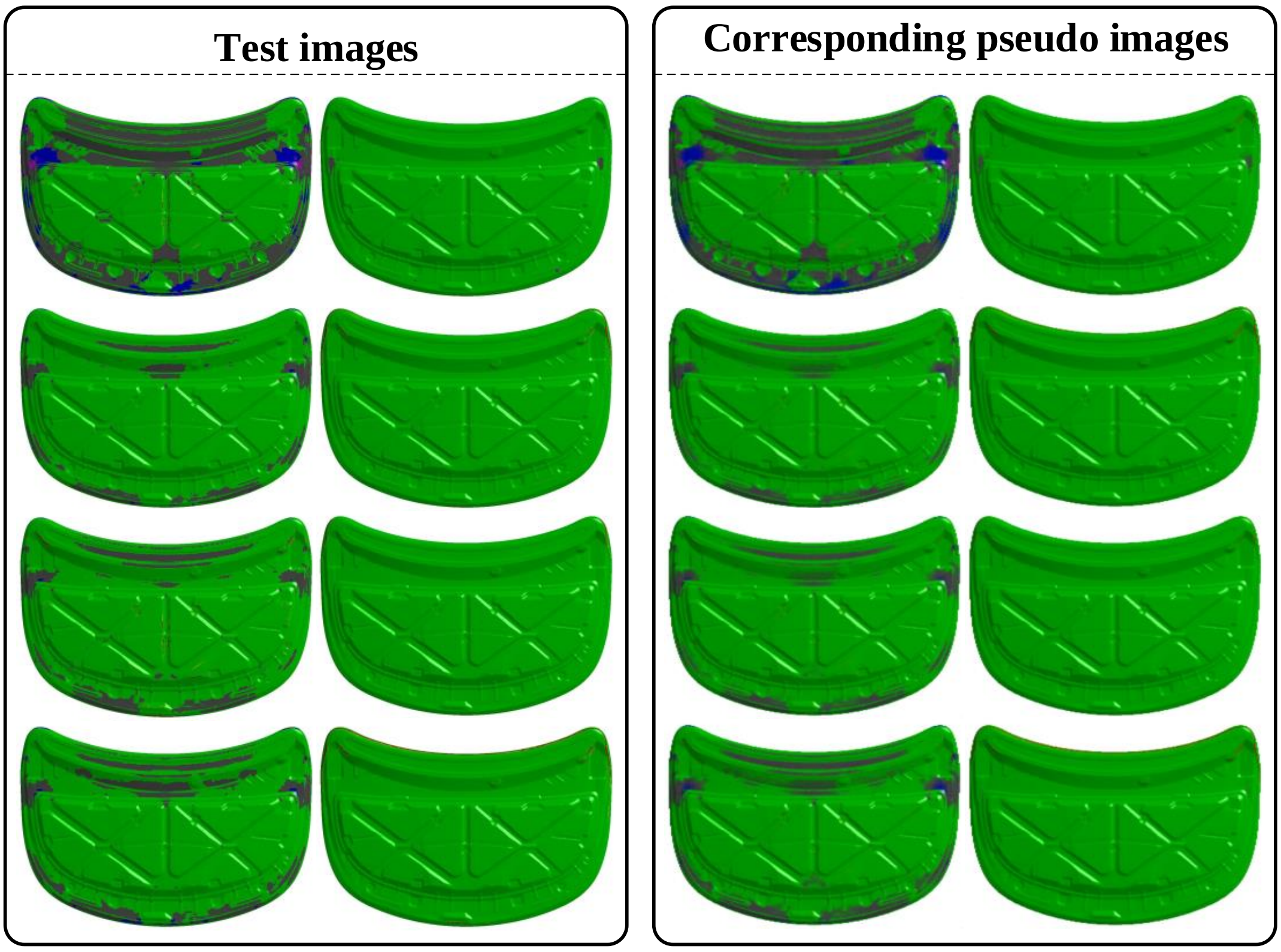

Fig. 10 The test images and corresponding pseudo images

The posterior distributions of the parameters obtained by comparing the observed coefficient vector $\mathbf{Zo}$ and simulated coefficient vector $\hat{\mathbf{Z}}$ with the ABC-NPMC sampling are shown in Fig. 11. The mean and STD of the posterior distribution are listed in Table 5. The VAE can reduce the original dimension to one-dimensional data, so the efficiency of ABC can be greatly improved by using reduced-dimensional data. Due to the image is processed as the objective function, the defect that the formability criteria in the traditional methods cannot fully reflect the forming quality is remedied. The working area is separated from the FLD by processing images, which prevents the interference of the non-working area for parameter inversion and makes the identification result more sufficient and effective. Simultaneously, the proposed method establishes an accurate LV-LSSVR surrogate model with a small number of samples, so that the simulated coefficient vectors in the ABC-NPMC sampling can be predicted with the established model to replace the time-consuming metal forming simulation. For the ABC-NPMC sampling, the tolerance $\varepsilon$ value decreases gradually with the increase of iteration steps, which ensures that the simulated coefficient vectors are closer to the observed coefficient vectors. Figure 12 shows the $\varepsilon$ value in the iteration process. It can be found that the final $\varepsilon$ value is very small, which means that the last reserved the simulated coefficient vector is infinitely close to the observed one. Therefore, the proposed method can effectively obtain the posterior distribution of the parameters, which is meaningful for uncertainty identification of parameters and dealing with complex sheet metal forming problem.

**Table 5** The mean and STD of the posterior distribution

| Material parameters | Mean | STD | Process parameters | Mean | STD |
|---|---|---|---|---|---|
| $E$ (*Gpa*) | 202.0251 | 5.5736 | $f_1$ | 0.1468 | 0.0058 |
| $u$ | 0.2964 | 0.0154 | $f_2$ | 0.1545 | 0.0075 |
| $K$ | 533.8378 | 7.5345 | $v$ (*mm*/*s*) | 4487.5259 | 218.7537 |
| $N$ | 0.3130 | 0.0038 | BHF (*Ton*) | 111.0795 | 2.8337 |
| $R_{00}$ | 1.9212 | 0.1554 | $F$ ($N$) | 140.3454 | 6.8865 |
| $R_{45}$ | 1.9009 | 0.1337 | | | |
| $R_{90}$ | 1.8994 | 0.1496 | | | |

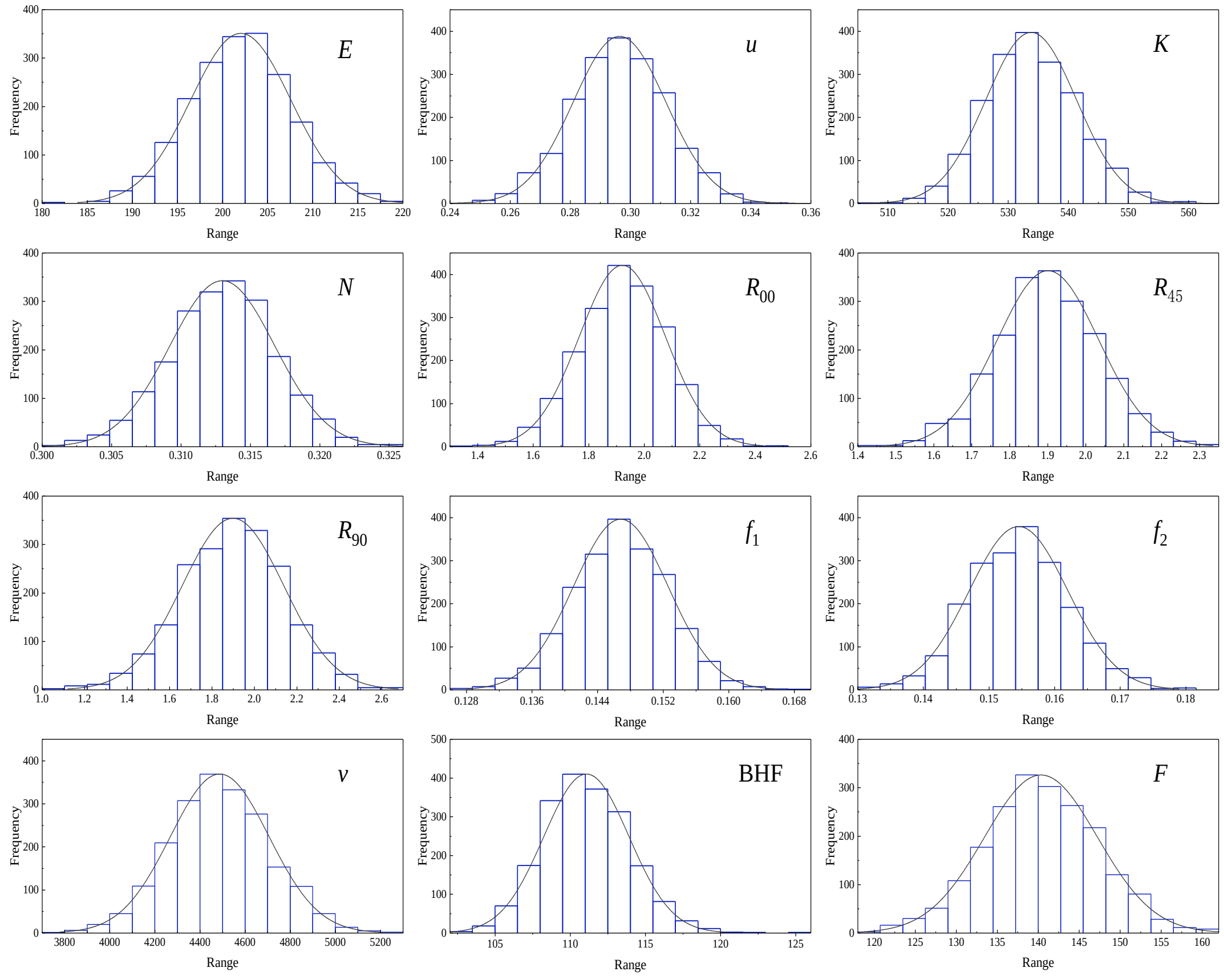

Fig. 11 The posterior distributions of uncertianty parameters

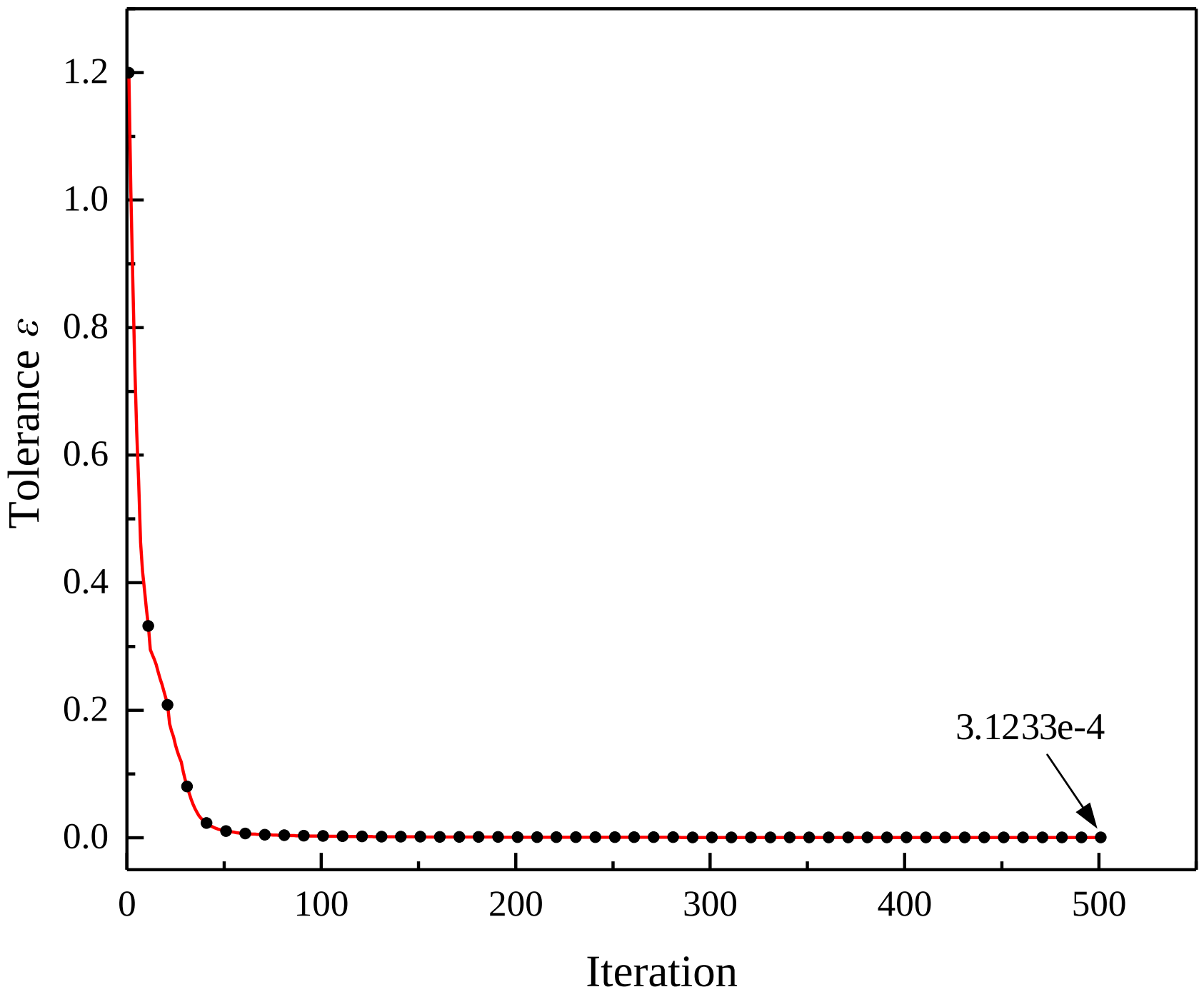

Fig. 12 The tolerance $\varepsilon$ value in the ABC-NPMC sampling

### 4.4 Verification of result

In this study, in order to verify the accuracy of the identification result, 50 samples are randomly selected from the posterior distribution and the FLDs are obtained by FE simulations. Meanwhile, the means of posterior distribution are also simulated to obtain a FLD. Figure 13 shows randomly selected 6 FLDs of previous 50 generated samples, the first of which is the results obtained with the mean of the posterior distribution. It can be found that there are still some cracks on the partial FLDs, but the area of cracks is relative small. Similarly, the wrinkles appear in the non-working areas of each FLD. For the FLD obtained with the mean of the posterior distribution, there is no wrinkle and crack in the working area.

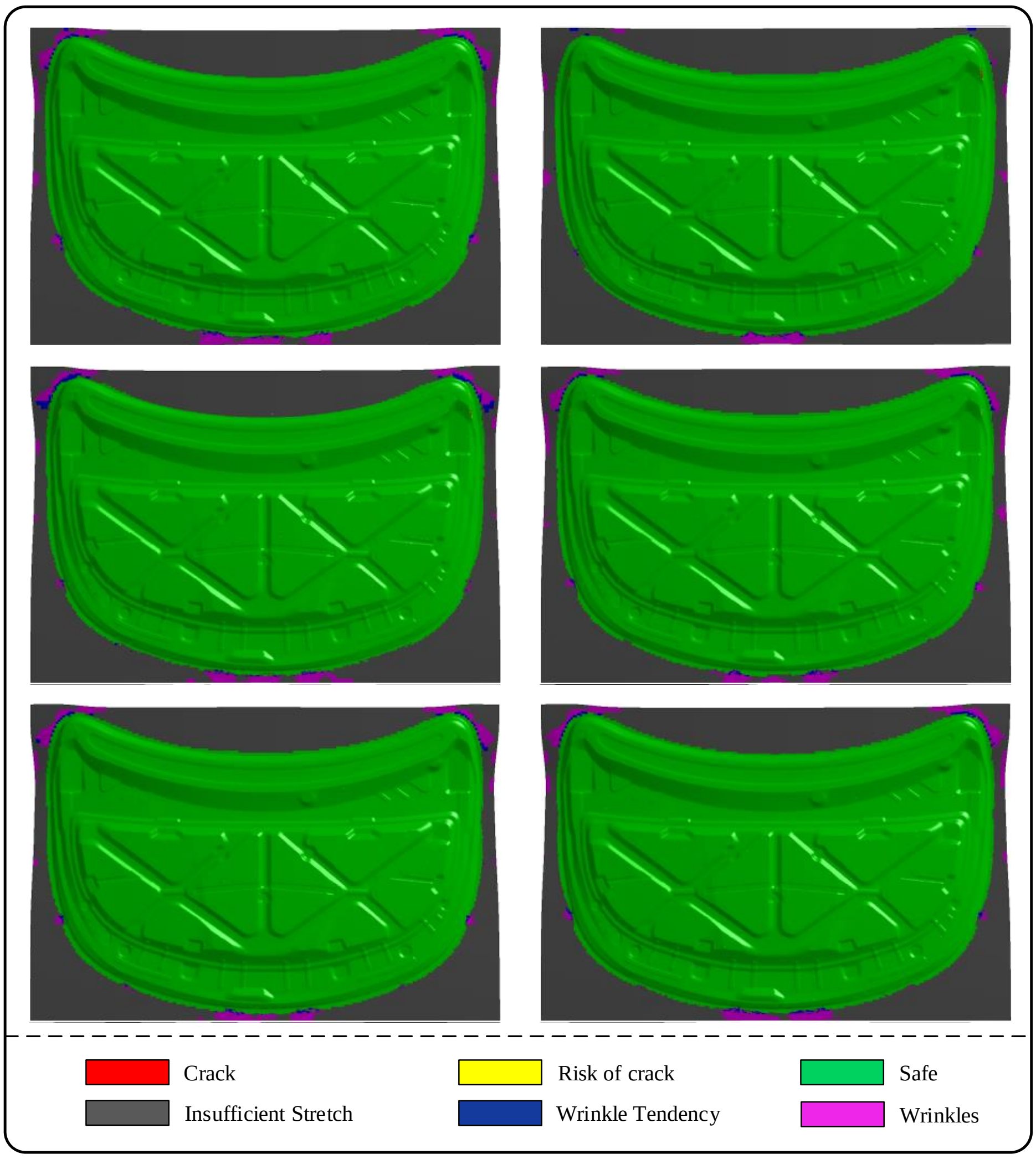

Fig. 13 The verification of result

The numbers of cracked elements and wrinkled elements in 51 FLDs are shown in Fig. 14. It can be found that some FLDs still have cracked elements, but the number of cracked elements is extremely small, which is acceptable. Although the number of the wrinkles elements is more than that of the cracked elements, they are all in the non-working area of the forming part, which can be considered as no influence on metal forming. Therefore, the parameter vector obtained by the proposed method is practical, which is of great significance for studying the metal forming properties. In addition, the means of posterior distribution can be used as the point estimation for the identified parameters, and the STD of the posterior probability distribution can be used as the uncertainty in the identification process.

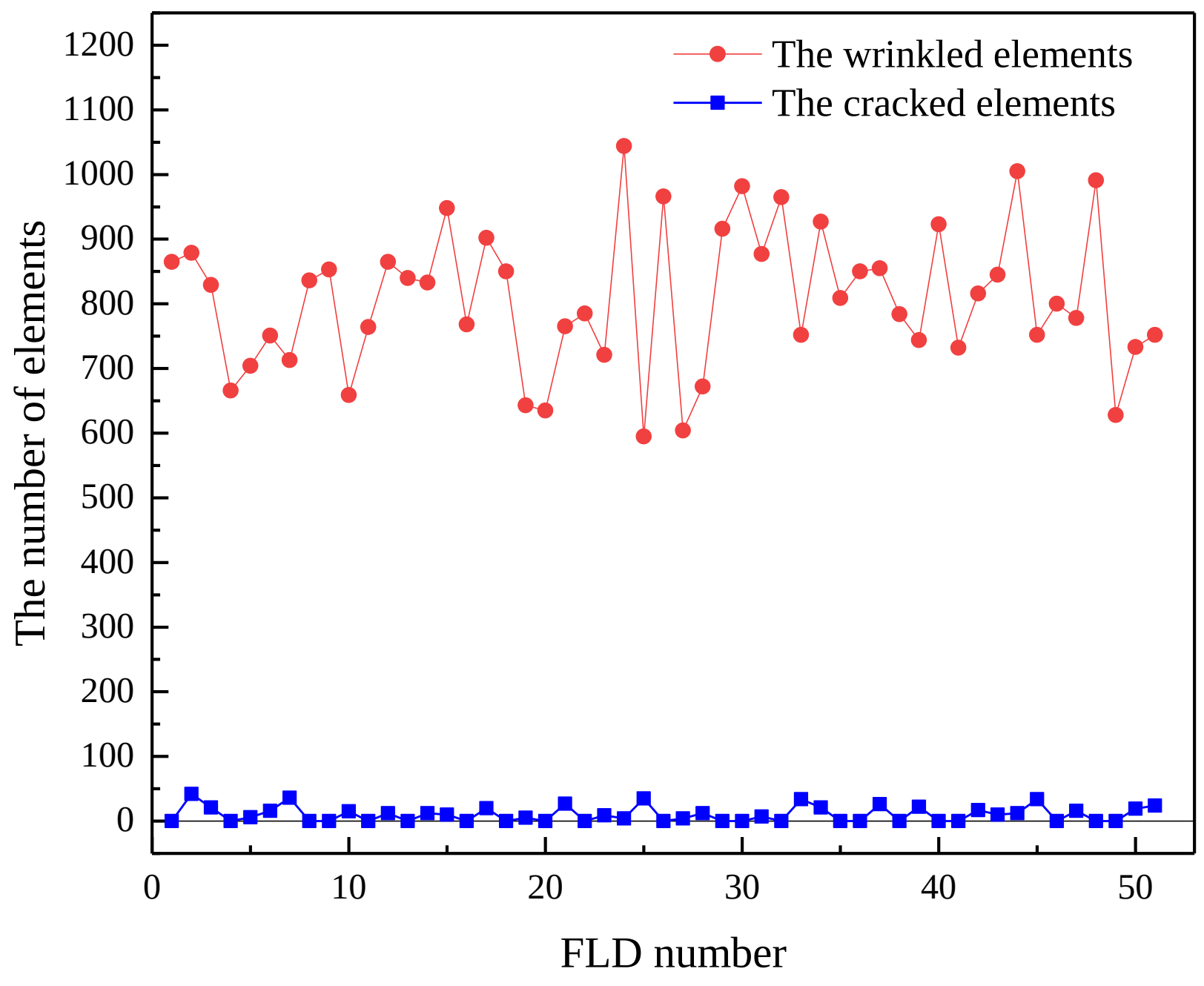

Fig. 14 The numbers of wrinkled and cracked elements

## 5 Conclusion

In this study, an novel image-assisted ABC parameter inverse method is proposed to identify design parameters for expensive evaluation problems. Compared with other popular methods, the uncertainties in the given problem are considered. Eventually, the probability distribution of the parameters should be identified by the ABC. Meanwhile, to save the computational cost, the VAE is used to reduce the dimension of the original problem. For a well-trained VAE model, its latent variables can be used instead of original ones as the summary statistics for the ABC. Then, the posterior distributions of parameters can be obtained by the ABC-NPMC sampling method. In addition, in order to further improve sampling efficiency, the LSSVR surrogate model is used instead of expensive evaluations to establish the relationship between parameter vectors and latent variables. Finally, uncertainty identifcaiton of material and process design varaibles in sheet metal forming is carried out by the suggested method succesfuly. Speically, a new forming criterion based on the image is used in this work. Gnenrally, the main contributions can be summarized as:

- Based on the advantages in feature extraction and sample reconstruction, the VAE is extended to construct the probability model in this study. Under such framework, the image can be mapped to low-dimensional latent space with minimal information loss, which can effectively compromise the information loss and computational cost of summary statistics in the ABC.
- In order to fully reflect the forming quality, the method of image recognition is directly used to define the formability criterion, which handles the drawback that the objective function cannot be accurately defined.
- In sheet metal forming, it is difficult to obtain the objective image directly by FE simulation due to its expensive computational cost. To solve this problem, the pesodu image should be generated from the evaluated cases and can be regarded as new samples in identicaitoon procedure.
- The VAE-based ABC method is feasibility and effective for identifying metal forming parameters. Furthermore, the proposed method can be applied to other similar parameter identification theoretically.

## Appendix A: LSSVR

For the given sample set $S=\{(\boldsymbol{\theta}_l,z_l),\boldsymbol{\theta}_l\in R^n,z_l\in R\}_{l=1}^m$, if its corresponding linear regression equation is

$$f(\boldsymbol{\theta})=\mathbf{w}^T\varphi(\boldsymbol{\theta})+b \tag{A.1}$$

where *m* denotes the sample size, $\mathbf{w}$ denotes the weight vector, $b$ denotes the model bias, and $\varphi(\boldsymbol{\theta})$ denotes the kernel function for solving the nonlinear problem. The squared error term and the equality constraint are introduced into the LSSVR, and Eq. (A.1) is rewritten as

$$\begin{aligned}&\min_{\mathbf{w},b,e} Q(\mathbf{w},b,e)=\frac{1}{2}\mathbf{w}^T\mathbf{w}+\frac{\gamma}{2}\sum_{i=1}^m e_i^2\\&s.t.\quad z_i=\mathbf{w}^{\mathbf{T}}\varphi(\boldsymbol{\theta}_i)+b+e_i\qquad i=1,2,\cdots,m\end{aligned} \tag{A.2}$$

where $e_i$ is the error between the estimated value and the true value of the *i*-th sample, and $\gamma$ is the regularization parameter. The Lagrange multiplier expression applied to Eq. (A.3):

$$L(\mathbf{w},b,e,\boldsymbol{\alpha})=Q(\mathbf{w},b,e)-\sum_{i=1}^m\alpha_i[\mathbf{w}^T\varphi(\boldsymbol{\theta}_i)+b+e_i-z_i] \tag{A.3}$$

where $\alpha_i$ is the Lagrange multiplier. By deriving the partial derivatives of $\mathbf{w},b,e$ and $\alpha$, the following equation is obtained.

$$\begin{cases}\dfrac{\partial L}{\partial \mathbf{w}}=0\Rightarrow\beta=\sum_{i=1}^m\alpha_i\varphi(\boldsymbol{\theta}_i)\\ \dfrac{\partial L}{\partial b}=0\Rightarrow\sum_{i=1}^m\alpha_i=0\\ \dfrac{\partial L}{\partial e_i}=0\Rightarrow\alpha_i=\gamma e_i,\quad i=1,\cdots,m\\ \dfrac{\partial L}{\partial \alpha_i}=0\Rightarrow\mathbf{w}^T\varphi(\boldsymbol{\theta}_i)+b+e_i-z_i=0,\ \ i=1,\cdots,m\end{cases} \tag{A.4}$$

Solving the equation above after eliminating $\mathbf{w}$ and $e$, one obtains the following linear expression

$$\begin{bmatrix} \mathbf{0} & \mathbf{I}^T \\ \mathbf{I} & \mathbf{\Lambda} + \frac{1}{\gamma}\mathbf{I} \end{bmatrix} \begin{bmatrix} b \\ \boldsymbol{\alpha} \end{bmatrix} = \begin{bmatrix} 0 \\ \mathbf{Z} \end{bmatrix} \tag{A.5}$$

where

$$\begin{cases} \mathbf{Z} = \left[z_1, z_2, \cdots, z_m\right]^T \\ \mathbf{I} = \left[1, 1, \cdots, 1\right]^T \\ \boldsymbol{\alpha} = \left[\alpha_1, \alpha_2, \cdots, \alpha_m\right]^T \\ \Lambda_{i,j} = \varphi(\boldsymbol{\theta}_i)^T \varphi(\boldsymbol{\theta}_j), \quad i, j = 1, 2, \cdots, m \end{cases} \tag{A.6}$$

The final LSSVR model for the function can be expressed

$$\hat{\mathbf{Z}} = \sum_{i=1}^{m} \alpha_i K(\boldsymbol{\theta}, \boldsymbol{\theta}_i) + b \tag{A.7}$$

## Acknowledgment

This work has been supported by Project of the Key Program of National Natural Science Foundation of China under the Grant Numbers 11572120 and 51621004, National Key Research and Development Program of China 2017YFB0203701, Key Projects of the Research Foundation of Education Bureau of Hunan Province (17A224).